\documentclass{iau} 
\usepackage{graphicx}
\usepackage{gensymb}
\usepackage{amsmath}

\title[LOFAR observations of the mode-switching pulsar B0943+10] 
{LOFAR observations of the mode-switching pulsar B0943+10}

\author[Anna V. Bilous]   
{Anna V. Bilous}

\affiliation{Anton Pannekoek Institute for Astronomy, 
    University of Amsterdam, Science Park 904, 
    1098 XH Amsterdam, The Netherlands, 
    email: {\tt A.Bilous@uva.nl}}

\pubyear{2017}
\volume{337}  
\jname{Pulsar Astrophysics -­ The Next 50 Years}
\editors{P. Weltevrede, B.B.P. Perera, L. Levin Preston \& S. Sanidas, eds.}

\begin{document}

\maketitle

\begin{abstract}
PSR B0943+10 is an old non-recycled pulsar which for decades has been mostly known for its rapid and 
spontaneous radio mode switching. Recently, \cite{Hermsen2013}  discovered correlated changes in 
the thermal X-ray emission from the polar cap, thus demonstrating that radio modes are not just a product 
of the local changes in the radio emission region, but a sign of some global magnetospheric transformation. 
At about the same time, owing to the commissioning of the new generation of low-frequency radio arrays,  
the broadband observations at the lowest edge of ionospheric transparency window became available. 
At these radio frequencies profile morphology and the single-pulse properties of PSR B0943+10's emission 
become very dynamic, providing details not only about the emission itself, but also about the conditions 
in the polar gap. Here, I will present the recent results of the LOFAR observations of PSR B0943+10 and 
discuss their contribution to the multiwavelength picture.
\end{abstract}

\firstsection 
\section{Introduction}
PSR~B0943+10 (hereafter B0943) is a relatively old, non-recycled pulsar. 
Once in a few hours, B0943 switches abruptly between so-called ``Bright'' and ``Quiet'' modes 
(hereafter B and Q). At radio frequencies, the pulsar is a few times brighter 
in B-mode and its single pulses form regular drifting patterns within the on-pulse phase window. 
With the start of Q-mode the drift disappears, the flux density drops and the shape of the average profile alters.
Mode transitions affect the properties of pulsar's high-energy emission as well \cite[(Hermsen \textit{et al.} 2013)]{Hermsen2013}. In X-rays, the pulsar 
is $\sim2.4$ times \emph{brighter} in the 
radio-Quiet mode and the emission has somewhat different spectral shape than in B-mode. X-ray emission from 
both modes has a pulsed thermal component coming from a hot polar cap \cite[(Mereghetti \textit{et al.} 2016)]{Mereghetti2016}.
The broadband nature of the mode-switching phenomenon may point to some global-scale magnetospheric 
transformation during the mode transition, making B0943 a useful laboratory
to study the complex interplay between various physical processes in the pulsar magnetosphere. However, the exact mechanism of mode 
switching remains unclear. 

\begin{figure}[h]
\begin{center}
 \includegraphics[trim=5mm 17mm 8mm 10mm,width=0.72\textwidth]{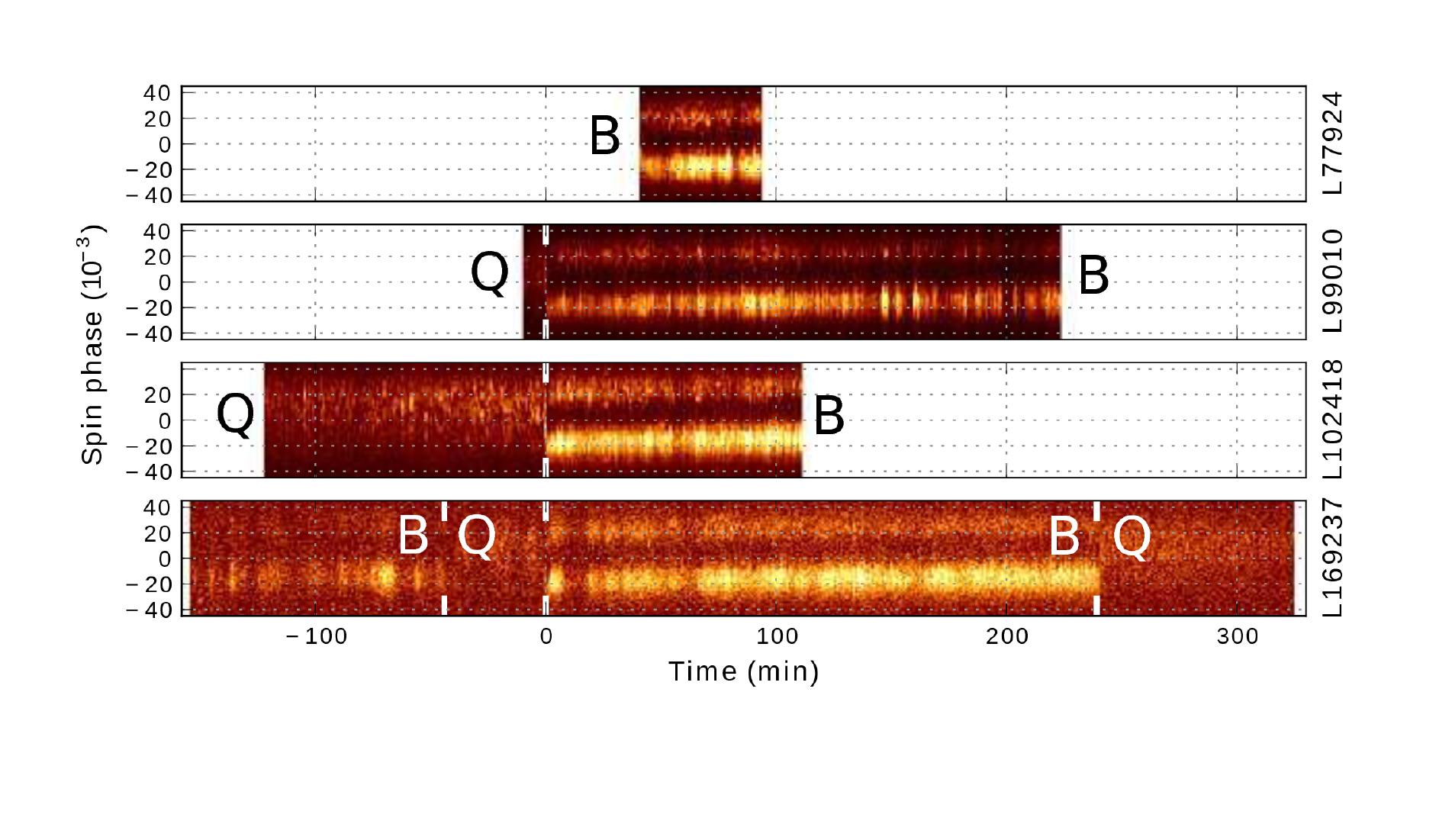}\includegraphics[width=0.28\textwidth]{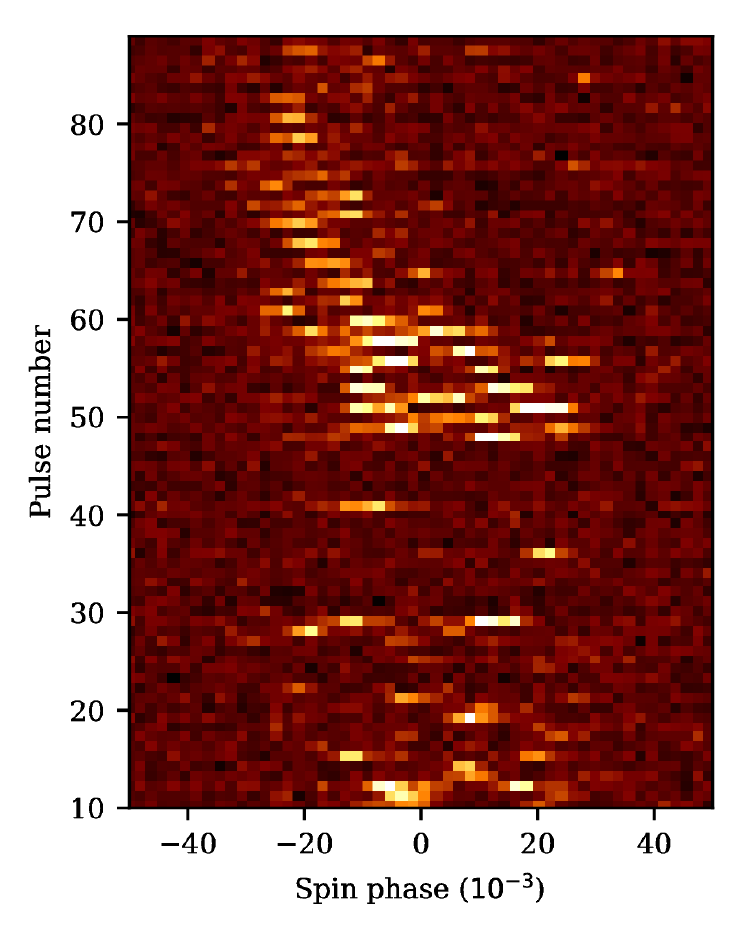}
 \caption{\textit{Left:} one-minute sub-integrations versus rotational phase for four
   observing sessions. Mode transitions are marked with white ticks. 
   \textit{Right:} example of single-pulse sequence around Q-to-B mode switch. The switch happens around pulse \#50.}
   \label{fig:data}
\end{center}
\end{figure}

Observing B0943 with the new generation of sensitive low-frequency radio telescopes with large fractional 
bandwidths may provide valuable input to solving the mode switching puzzle. At the frequencies below 200\,MHz 
the average profile morphology starts evolving rapidly both in B- and Q-modes. Within the framework of 
radius-to-frequency mapping theory (RFM), it means that below 200\,MHz a relatively larger range 
of emission heights can be observed at once. At the same time, the pulsar is quite bright in this frequency 
range, so it is possible to track the changes in the average profile shape on the time scales of few minutes 
and also to observe the individual pulses directly. 

In 2012--2013, B0943 was observed with the LOFAR (LOw FRequency ARray) core stations, with several full emission modes being
recorded in the broad 25--80\,MHz frequency range (Fig.~\ref{fig:data}, left). Here I present a summary of the 
results, originally published in \cite{Bilous2014} and \cite{Bilous2017}.

\section{Lessons learned}

\underline{Revisiting geometry and the number of sparks in the carousel.} 
Based on the analysis of the frequency-dependent separation between B-mode average profile components at the frequencies below 300\,MHz
and the brief presence of a 37-period modulation in the amplitudes of drifting subpulses, 
\cite{Deshpande2001} concluded that B0943 is an almost aligned rotator, 
with the inclination angle $\alpha$ within $12^\circ$--$16^\circ$, and the line-of-sight LOS passing 
between the spin and magnetic axes (yielding the impact angle $\beta$ of about $4^\circ$--$6^\circ$). 
The time-integrated part of the geometry derivation was successfully reproduced using LOFAR observations. 
However, it must be stressed that the exact values of geometry angles 
depend heavily on the assumed opening angle of the emission cone at 1\,GHz, $\rho_\mathrm{1\,GHz}$, namely $\alpha\approx 3\rho_\mathrm{1\,GHz}$ 
and $\beta\approx \rho_\mathrm{1\,GHz}$. 
Taking the distribution of $\rho_\mathrm{1\,GHz}$ from the work of \cite{Mitra1999}
yields an $\alpha$ of $5^\circ$--$25^\circ$ and,
in principle, the current geometry derivation scheme allows for almost any value of $\alpha$.
The weaker constraints on geometry angles have to be taken into account while explaining the unexpectedly large fractional amplitude
of the thermal X-ray emission from B0943's polar caps \cite[(Mereghetti \textit{et al.} 2016)]{Mereghetti2016}. Another interesting 
constraint on B0943's geometry may come from the direct modelling of the light curve of X-ray pulsations.

In the work of \cite{Deshpande2001}, the orientation of the line-of-sight vector with respect to magnetic and spin axes
was set by the presence of periodic amplitude modulation of drifting subpulses (so-called ``tertiary modulation''). 
So far, only two more brief detections of tertiary modulation are known, both of them above 300\,MHz \cite[(Backus \etal\ 2010)]{Backus2010}.
Despite the dedicated searches and relatively large volume of B-mode observations, no sign of tertiary modulation was detected
in the LOFAR data. 
Since tertiary modulation is a vital argument for constraining several other
important system parameters: (e.g. the degree of subpulse drift aliasing, the angular velocity of the carousel, and thus, 
the gradient of the accelerating potential
in the polar gap), obtaining strong independent confirmation of it is of utmost importance.

\begin{figure}[h]
\begin{center}
 \includegraphics[width=0.45\textwidth]{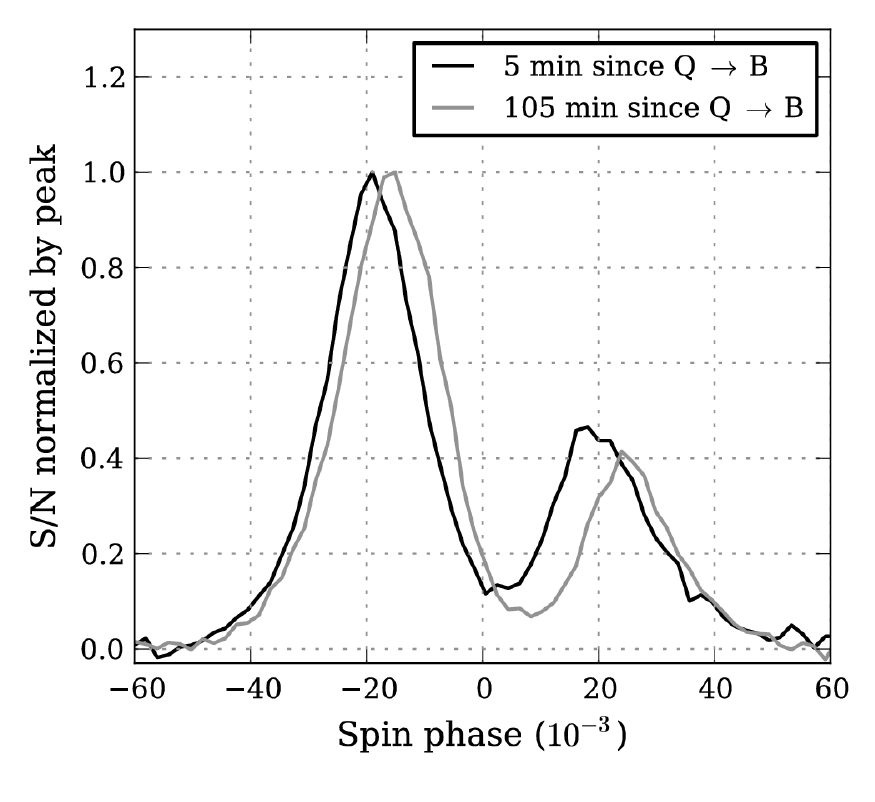}\includegraphics[width=0.45\textwidth]{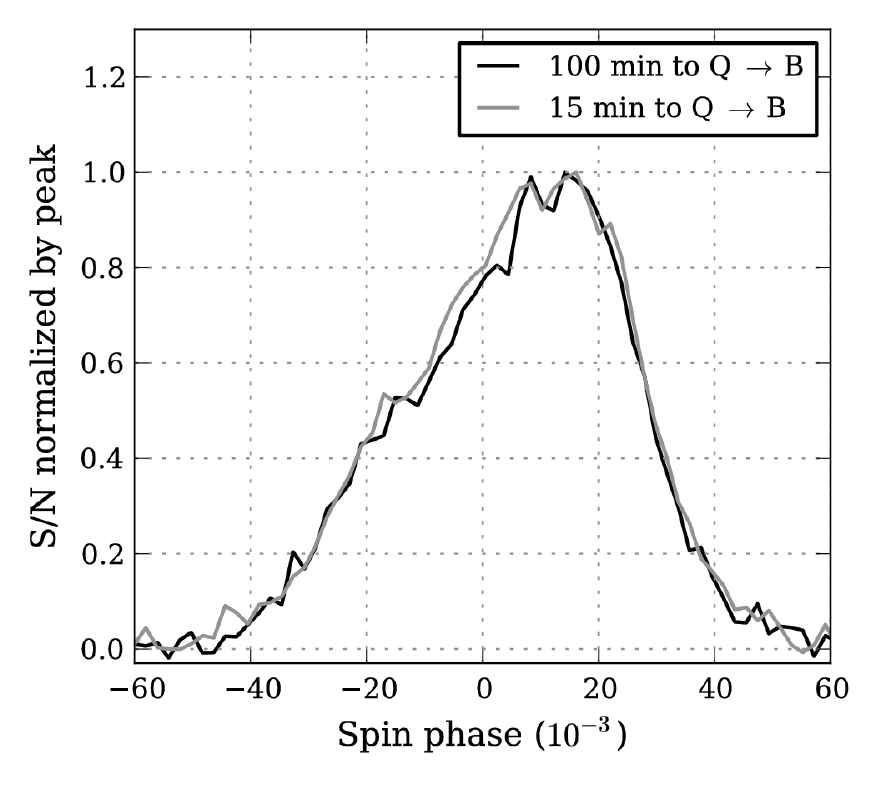} 
 \caption{ Two average profiles of the
Q-mode (at the start of observations and just before the mode transition)
and the B-mode (right after mode transition and at the end of the
observation) for the observing session L102418. Noticeably, 
in addition to changes in relative intensity of the
components and their width, the profile in B-mode shifts as a whole towards later spin phases.}
   \label{fig:delay}
\end{center}
\end{figure}

\underline{Emission height and fiducial longitude.} 
The midpoint between profile components in the LBA B-mode data appeared to
follow the $\nu^{-2}$ dispersion law down to 25\,MHz, 
allowing us to measure DM independently of the profile evolution. 
This implies that the position of midpoint is frequency-independent and, 
thus, is well suited for the role of fiducial longitude, $\phi_0$.
The absence of aberration and retardation  signatures in the profile 
midpoint placed the emission region (at a height $r$) much 
closer to the stellar surface than to the light cylinder ($r_\mathrm{LC}$): 
$r/r_\mathrm{LC}<0.06$. 

\underline{Systematic delay of fiducial longitude}. 
It is well known that in the B-mode both the average profile and single
pulses evolve systematically during a mode instance.  
LOFAR observations discover one more feature of the B-mode profile evolution: 
the frequency-independent midpoint between B-mode profile components is systematically shifting towards
later spin phases with the time from the start of the mode.  
The observed lag 
asymptotically changes towards a stable value of 4\,ms, 
much too high to be due to a changing spin-down rate. Similar results have been obtained by \cite{Suleymanova2014}
around 100\,MHz.
At the B-to-Q transition the profile midpoint jumps back to the earlier spin phase and remains
constant throughout Q-mode until the next Q-to-B transition (Fig.~\ref{fig:delay}).
An interesting explanation of the midpoint lag involves the variation of accelerating potential between the surface of the pulsar and 
the start of the plasma-filled magnetosphere above the polar gap, which is responsible for the 
departure of corotation in the plasma-filled magnetosphere \cite[(Melrose \& Yuen 2014)]{Melrose2014}. This explanation
connects the observed midpoint lag to the gradually evolving subpulse drift rate, which
is determined by the gradient of potential in orthogonal
direction -- across the field lines in the polar cap. 

\begin{figure}[h]
\begin{center}
\includegraphics[width=0.33\textwidth]{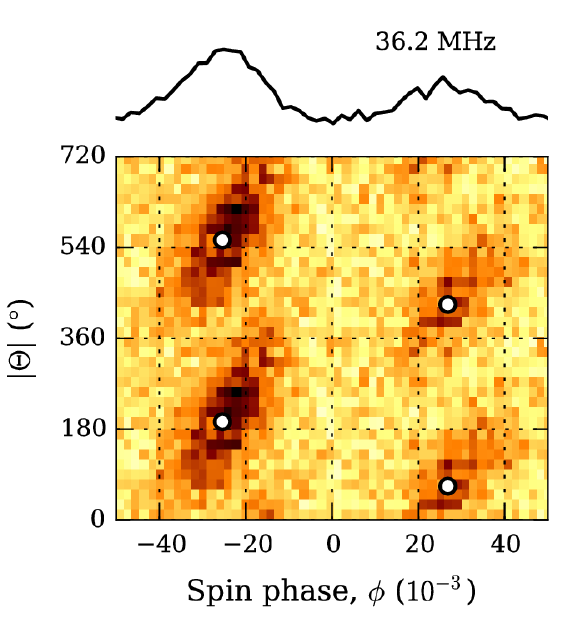}\includegraphics[width=0.33\textwidth]{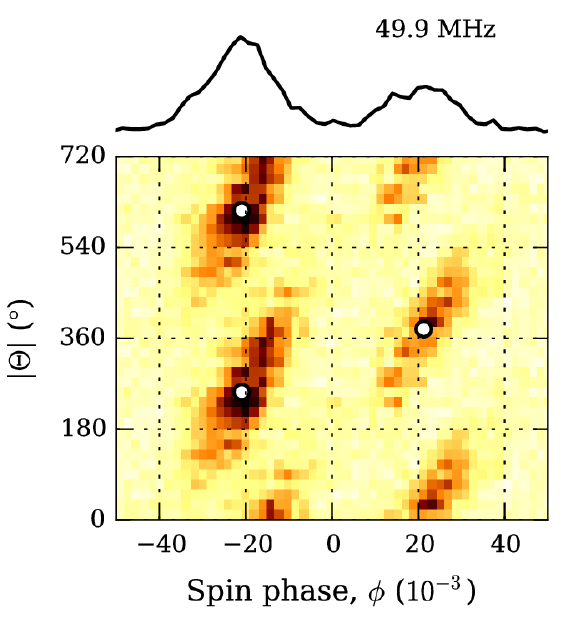}\includegraphics[width=0.33\textwidth]{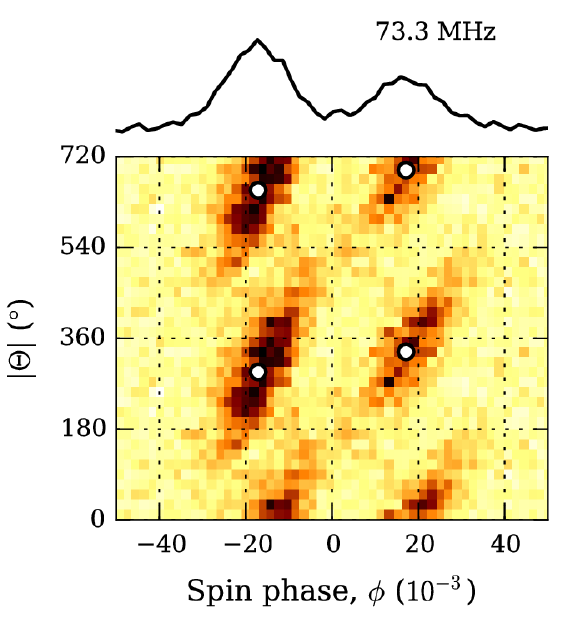}
 \caption{An example of one of 512-pulse stacks folded modulo drift period $P_3$ for three radio frequencies within LBA band. 
 The modfold is repeated twice along the Y-axis to mitigate wrapping, and the corresponding average profile is plotted on 
 top. White circles mark the centres of driftbands obtained from fitting two-dimensional tilted Gaussians. As the radio 
 frequency increases, the driftbands move towards each other both in $\phi$ (reflecting the behaviour of the average 
 profile components) and  in drift phase $\Theta$.}
   \label{fig:driftbands}
\end{center}
\end{figure}

\begin{figure}[h]
\begin{center}
 \includegraphics[width=0.65\textwidth]{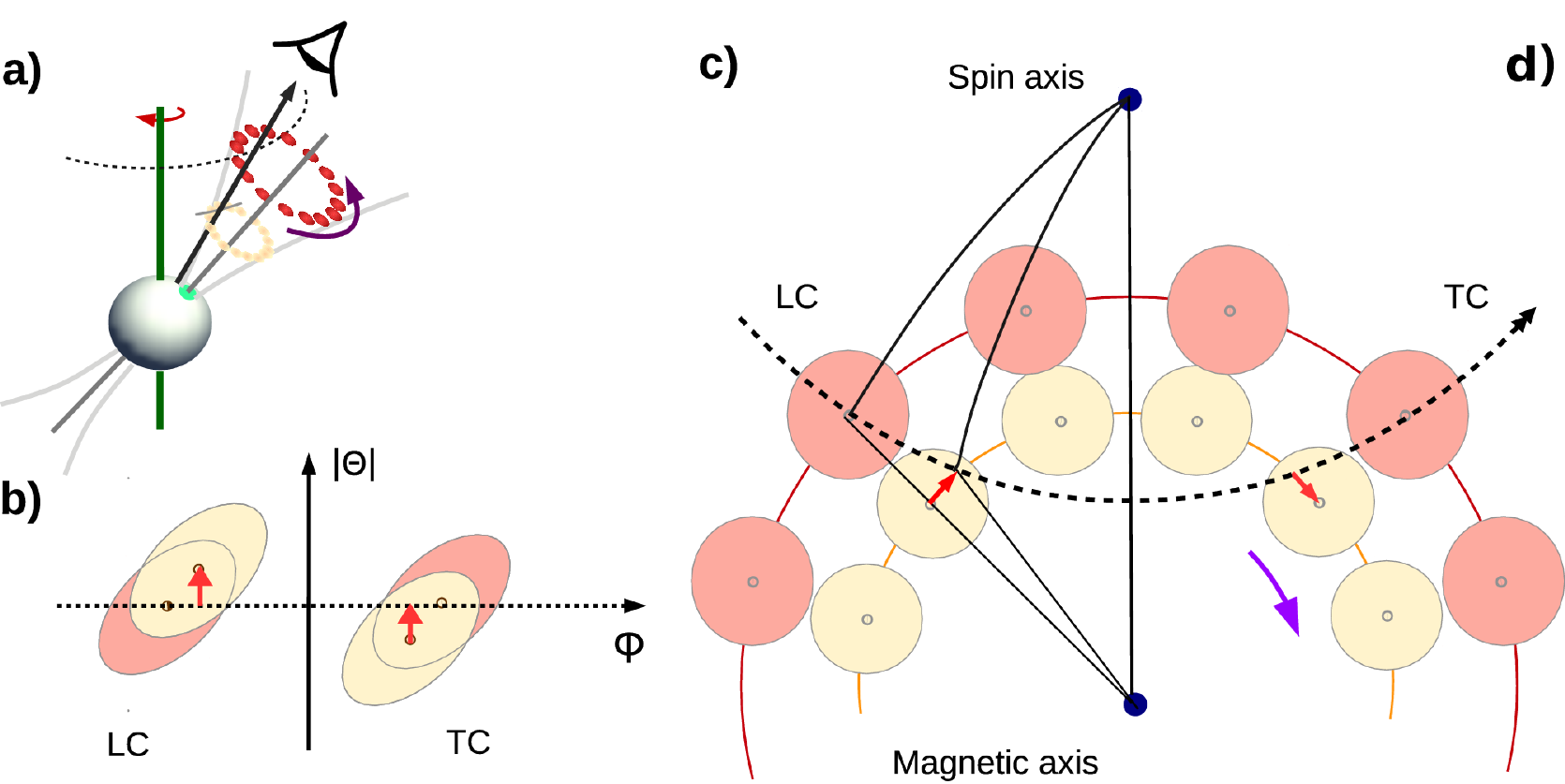}\includegraphics[width=0.35\textwidth]{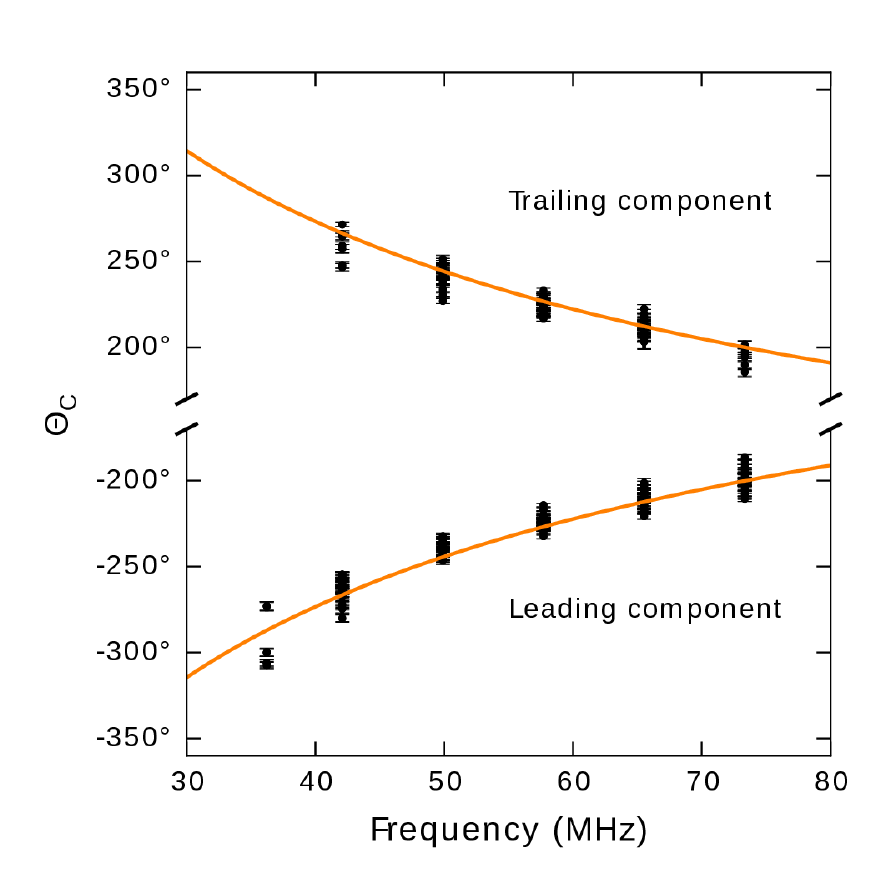} 
 \caption{Cartoon model of the drift track evolution: \textit{(a)} general view on the system (inside traverse geometry,
 pulsar is rotating clockwise) with individual sparks at two different radio frequencies, 
 $\nu_\mathrm{lo}$ (darker) and $\nu_\mathrm{hi}$ (lighter). 
 The modfolds at these two frequencies are shown in \textit{(b)}. The carousel 
 configuration corresponding to the moment marked with dotted horizontal line on \textit{(b)} is shown in \textit{(c)}. 
 The arrow marks the direction of carousel rotation. 
 For this moment, the LOS sweeps  through the centres of sparks at $\nu_\mathrm{lo}$ and misses the spark centres at $\nu_\mathrm{hi}$. 
 For the leading component (LC), the  rotating carousel needs more time to bring the centre of spark to the LOS. Drift phase $\Theta$ is directly proportional 
 to time, thus  $|\Theta_\mathrm{LC}(\nu_\mathrm{hi})|>|\Theta_\mathrm{LC}(\nu_\mathrm{lo})|$. 
 \textit{(d):} black points mark drift phase $\Theta_\mathrm{C}$ versus radio frequency for the leading and 
trailing drift tracks. The orange lines show the phase delay of the rotating carousel calculated according to Eq.~\ref{eq:Theta}.}
   \label{fig:cartoon}
\end{center}
\end{figure}

\underline{Frequency-dependent drift phase delay.} A common technique of visualizing the tracks of drifting subpulses is folding 
them modulo drift period $P_3$. In 2013, Hassall \textit{et al.} discovered that the centres of driftbands for PSR B0809+74 evolve with 
radio frequency. 
A similar effect was observed for B0943 (Fig.~\ref{fig:driftbands}). 
It can be shown that this delay can be quantitatively modelled within the rotating carousel model in the RFM convention (see also Fig.~\ref{fig:cartoon}):
\begin{equation}
\label{eq:Theta}
 \Theta_\mathrm{C}(\nu) =  -N\,\mathrm{sgn}\,\beta\frac{\sin(\phi_\mathrm{C}-\phi_0)\sin\zeta}{\cos\zeta\sin\alpha-\cos(\phi_\mathrm{C}-\phi_0)\sin\zeta\cos\alpha} +(\phi_\mathrm{C}-\phi_0)\frac{P_1}{P_3} + \Theta_0, 
\end{equation}
where subscript C marks the centre of driftband, 
$\Theta_0= \Theta_\mathrm{C}(\nu=\infty)$, $N$ is the number of sparks in the carousel, $P_1$ is the pulsar period, $P_3$
is the subpulse drift period, and $\zeta=\alpha+\beta$. Similar quantitative explanation can be also applied to PSR B0809+74.

\end{document}